# The temporal evolution of exposed water ice-rich areas on the surface of 67P/Churyumov-Gerasimenko: spectral analysis.

A. Raponi[1], M. Ciarniello[1], F. Capaccioni[1], G. Filacchione[1], F. Tosi[1], M. C. De Sanctis[1], M.T. Capria[1], M. A. Barucci[2], A. Longobardo[1], E. Palomba[1], D. Kappel[3], G. Arnold[3], S. Mottola[3], B. Rousseau[2], E. Quirico[4], G. Rinaldi[1], S. Erard[2], D. Bockelee-Morvan[2], C. Leyrat[2]

[1]*IAPS-INAF, via del Fosso del Cavaliere 100, Rome 00133, Italy*
[2]*LESIA, Observatoire de Paris/CNRS/Université Pierre et Marie Curie/Université Paris-Diderot, Meudon, France*
[3]*Institute for Planetary Research, Deutsches Zentrum fur Luft- und Raumfahrt (DLR), Berlin, Germany*
[4]*Université Grenoble Alpes, CNRS, IPAG, Grenoble, France*

**Abstract.** Water ice-rich patches have been detected on the surface of comet 67P/Churyumov-Gerasimenko by the VIRTIS hyperspectral imager on-board the Rosetta spacecraft, since the orbital insertion in late August 2014. Among those, three icy patches have been selected, and VIRTIS data are used to analyse their properties and their temporal evolution while the comet was moving towards the Sun. We performed an extensive analysis of the spectral parameters, and we applied the Hapke radiative transfer model to retrieve the abundance and grain size of water ice, as well as the mixing modalities of water ice and dark terrains on the three selected water ice rich areas.

Study of the spatial distribution of the spectral parameters within the ice-rich patches has revealed that water ice follows different patterns associated to a bimodal distribution of the grains: ~50 μm sized and ~2000 μm sized.

In all three cases, after the first detections at about 3.5 AU heliocentric distance, the spatial extension and intensity of the water ice spectral features increased, it reached a maximum after 60-100 days at about 3.0 AU, and was followed by an approximately equally timed decrease and disappearance at about ~2.2 AU, before perihelion. The behaviour of the analysed patches can be assimilated to a seasonal cycle. In addition we found evidence of short-term variability associated to a diurnal water cycle.

The similar lifecycle of the three icy regions indicates that water ice is uniformly distributed in the subsurface layers, and no large water ice reservoirs are present.

**Introduction**

Comets are primordial bodies, which played a key role in the early phases of solar system formation contributing to the accretion of the giant planets as well as potentially contributing volatiles to the terrestrial planets (A'Hearn et al., 2011a). Water is the most abundant volatile species in comets as demonstrated by long-term studies of cometary comae (e.g. Bockelée-Morvan and Rickman, 1997; Bockelée-Morvan et al., 2004; Feaga et al., 2007; Mumma and Charnley, 2011).

Water was assembled, presumably, as amorphous ice within the cometesimal, as a consequence of the very low formation temperatures of the icy grains. Evolution of the surface, and of the ice therein contained, of Jupiter Family Comets (JFC) is driven by their periodic passages close to the Sun. The surface is modified by sublimation of volatiles leading to surface erosion, which causes material's loss up to tens of meters at each perihelion passage, and by heating, which alters the structure of the upper layers of the nucleus to a depth determined by the thermal skin depth. In summary, at each periodic passage within the inner Solar System, primordial water ice is subjected to evolutionary processes such as phase transition, sublimation, re-condensation, sintering. Amorphous ice is likely not observable directly at the surface of a cometary nucleus at close distances to the Sun as it undergoes a phase transition with an activation temperature of 120 K, which transforms amorphous ice into crystalline ice.

The study of the state of water ice (crystalline Vs. amorphous, grain size, presence of clathrates, etc.) at the surface of a cometary nucleus can provide some insight into the processes that have led to its evolution.

The properties investigated by the present work are: abundance, grain size, and mixing modalities of the ice with the refractory dark material of comet. Hereafter, the abundance is intended as the total cross section of the icy grains as a fraction of the area subtended by the pixel. The grain size is the physical diameter of the spherical grain, as in Hapke (2012). The mixing modalities are assumed to be either intimate or areal. In the intimate mixing the grains of the ice and dark terrain are in contact each other. In the areal mixing the ice is present in form of patches on the surface.

A first insight on the properties cometary water ice was provided by the in situ observations carried out by the Deep Impact mission to 9P/Tempel 1 (A'Hearn et al., 2005) and its investigation of 103P/Hartley 2 during the extended mission (A'Hearn et al., 2011b). However, a real breakthrough was represented by the Rosetta mission to the comet 67P/Churyumov-Gerasimenko (hereafter referred to as 67P) which carried out the most complete set of observations, performing the first escort and soft landing on a comet.

The more diffuse form in which the water ice has been detected in different comets is a population of pure and very fine grains (~1 μm in diameter) of crystalline ice:
- Ground observation of the large and infrequent outbursts of comet 17P/Holmes revealed pure and fine water ice grains (Yang et al., 2009);
- The material excavated and ejected from 10 to 20-m depths from comet 9P/Tempel 1 by the impactor released by the Deep Impact spacecraft has been modelled with 1 μm pure water ice grains (Sunshine et al., 2007);
- The innermost coma of the hyperactive comet 103P/Hartley 2 has been found to be populated by 1 μm sized grains, which are components of most massive aggregates dragged out continuously by gaseous $CO_2$ (Protopapa et al., 2014);
- Observations, by the VIRTIS instrument (Coradini et al., 2007) on-board the Rosetta spacecraft, of the neck region on the surface of 67P revealed fine grains (1-2 μm) intimately mixed with the predominant dark terrain, which has been suggested to be the result of a diurnal sublimation-recondensation cycle (De Sanctis et al., 2015);
- An increasing exposure of fine grains (1-2 μm) of water ice on 67P observed after the gradual removal of the surficial dust layer by the gaseous activity, before perihelion (Filacchione et al., 2016b; Ciarniello et al., submitted).

However, there is at least another form of cometary water ice: in some specific and small areas on the surface of comets it has been observed in form of patches with coarser grain sizes (> 30 μm). Specifically:
- On the surface of comet Tempel 1 water ice has been detected by the HRI spectrometer onboard the Deep Impact spacecraft in three regions covering a total area of 0.5 km$^2$. The icy areas have been modeled with a few percent of water ice constituted by 30 μm sized grains, in areal mixture with the dark terrain of the comet (Sunshine et al., 2006), and by up to 1%, with 30 or 70 μm sized grains, in areal or intimate mixture, respectively (Raponi et al., 2013);
- On the southern hemisphere of 67P icy patches have been detected in correspondence of two debris falls. Spectral modelling has revealed a bimodal distribution of grains: ~50 μm sized and ~2000 μm sized, respectively in intimate and areal mixture with the dark refractory material of the comet (Filacchione et al., 2016a);
- Other dozen small patches or cluster of spots spread over the entire surface of 67P, and covering up to few hundred square meters each one, have been detected by OSIRIS (Keller et al., 2007) and VIRTIS instruments on board the Rosetta spacecraft (Pommerol et al., 2015, Barucci et al., 2016).

In order to have an overall view of the physical state and processes of water ice on comets, intrinsic differences between the comets observed, in terms of composition, activity and surface morphology must be taken into account. The evolution of comets is then subjected to seasonal effects, related to the total solar input received by the comets along an entire orbit, and to short term processes, attributable to the instantaneous solar input.

The analysis of the continuous evolution in time of the water ice on the surface of comets was not possible before the Rosetta mission, because all the previous space missions performed scientific measurements during only relatively short flybys. It is worth noting that one example existed, before Rosetta, of a long term analysis of a surface of a comet thanks to Stardust-NEXT which on 14 February 2011 flew past 9P/Tempel 1 showing evident differences in some geological features with respect to the first flyby of the Deep Impact spacecraft on 4 July 2005 (Veverka, 2012).

The Rosetta mission has provided the first ever escort of a comet along its orbit around the Sun, allowing us to follow the orbital evolution of the surface during the perihelion passage.

The aim of the present work is the analysis of the water ice patches on the surface of 67P showing the presence of coarser grains, which have been suggested as more stable in time (Filacchione et al., 2016a; Barucci et al., 2016). In particular, here we focus on the spectral parameters, the physical properties, and their evolution in time, using the hyperspectral data acquired by the VIRTIS instrument on-board Rosetta.

The VIRTIS (Visible InfraRed and Thermal Imaging Spectrometer) instrument is described in detail by Coradini et al. (2007); summarizing, VIRTIS is composed of a high spectral resolution channel (VIRTIS-H) and a hyperspectral mapper (VIRTIS-M) covering the spectral range 0.22 - 1.04 µm (VIRTIS-M-VIS) and 0.95 - 5.06 µm (VIRTIS-M-IR) with two detectors both with 432 spectral bands. The present work takes advantage of the data acquired simultaneously by VIS and IR channels of the VIRTIS-M instrument.

**Dataset description**

We have taken into account all the water ice patches discussed by Filacchione et al. (2016a) and by Barucci et al. (2016). We run a search in the full VIRTIS observations database to retrieve all observations covering the regions where icy patches were observed. The acquisitions with unfavourable viewing geometry (incidence or emission angle > 70°), or with limited spatial coverage were discarded. The patches with a limited temporal coverage were also discarded. In addition, all pixels in shadow were excluded from the analysis.

Our final selection included three large patches: the "BAPs" (bright albedo patches) discussed by Filacchione et al. (2015): BAP1 (longitude: 118°E, latitude: 13°N), BAP2 (longitude: 180°E, latitude: -4°N), and the "SPOT 6" (longitude: 72°E, latitude: 3°N) discussed by Barucci et al. (2016). The surface portions that are related to patches are covered by ~50 - ~5000 pixels respectively from the least to the highest resolved acquisitions.

The measured spectra have been cleaned from artifacts and spikes (Raponi, 2014). The VIS and IR channels have been co-registered to bridge the spectra of the two channels.

The VIRTIS-M dataset considered in this work is summarized in Tables 1, 2, and 3.

| Spacecraft clock time | Resolution (m/px) | Local solar time (h) | Average incidence angle (°) | Average emission angle (°) | Average phase angle (°) | Days from 2 Sept 2014 (first acquisition) | Helio centric distance (AU) |
|---|---|---|---|---|---|---|---|
| 368245714 | 12.7 | 10.0025 | 56.4 | 48.1 | 38.8 | 0.000 | 3.443 |
| 368291071 | 14.0 | 9.3979 | 60.7 | 29.3 | 45.2 | 0.525 | 3.440 |
| 369364114 | 7.1 | 10.1517 | 58.4 | 44.9 | 67.5 | 12.944 | 3.365 |
| 371998843 | 2.2 | 9.3045 | 58.3 | 55.0 | 95.0 | 43.439 | 3.174 |
| 373202012 | 4.4 | 8.0281 | 67.0 | 40.5 | 96.2 | 57.365 | 3.084 |
| 373248812 | 5.8 | 8.9563 | 61.1 | 42.0 | 94.7 | 57.906 | 3.080 |
| 375524971 | 7.3 | 8.2379 | 63.7 | 44.7 | 97.8 | 84.251 | 2.904 |
| 375613471 | 7.2 | 7.6880 | 67.6 | 53.2 | 97.8 | 85.275 | 2.897 |
| 383518966 | 20.4 | 7.9924 | 53.0 | 45.0 | 51.7 | 176.77 | 2.230 |

Table 1. Dataset selected for BAP 1: The first column indicates the spacecraft clock time which is also the name of the acquisition. The diverse resolutions are due to the different distance of the spacecraft from the comet surface at the time of the acquisitions. The column "local solar time" is defined according to the longitude position, and it is in hour unit, being the comet rotation period divided in 24 hours.

| Spacecraft clock time | Resolution (m/px) | Local solar time (h) | Average incidence angle (°) | Average emission angle (°) | Average phase angle (°) | Days from 25 Aug 2014 (first acquisition) | Helio centric distance (AU) |
|---|---|---|---|---|---|---|---|
| 367617963 | 12.5 | 12.9488 | 62.5 | 56.0 | 37.0 | 0.000 | 3.485 |
| 367621623 | 12.5 | 14.4624 | 58.0 | 40.8 | 36.7 | 0.043 | 3.486 |
| 368299474 | 15.0 | 17.6636 | 64.2 | 60.6 | 47.4 | 7.888 | 3.439 |
| 369369214 | 7.7 | 17.0423 | 60.2 | 66.2 | 67.9 | 20.269 | 3.364 |
| 377184571 | 5.0 | 16.0833 | 53.7 | 55.8 | 92.2 | 110.726 | 2.770 |
| 384381667 | 19.2 | 19.7250 | 69.2 | 34.0 | 47.3 | 194.026 | 2.153 |

**Table 2**. Same as Table 1 but for BAP 2.

| Spacecraft clock time | Resolution (m/px) | Local solar time (h) | Average incidence angle (°) | Average emission angle (°) | Average phase angle (°) | Days from 24 Aug 2014 (first acquisition) | Helio centric distance (AU) |
|---|---|---|---|---|---|---|---|
| 367589075 | 12.9 | 13.9854 | 50.7 | 26.5 | 39.1 | 0.000 | 3.488 |
| 367592759 | 12.9 | 15.3450 | 55.3 | 33.8 | 38.8 | 0.043 | 3.488 |
| 369374074 | 7.1 | 11.9664 | 57.4 | 31.5 | 68.9 | 20.66 | 3.364 |
| 373532092 | 8.2 | 14.2728 | 55.9 | 56.9 | 108.8 | 68.785 | 3.059 |
| 376302211 | 5.9 | 14.9415 | 53.3 | 49.3 | 91.0 | 100.846 | 2.842 |
| 383489908 | 19.9 | 12.8080 | 31.9 | 52.5 | 51.9 | 184.04 | 2.233 |

**Table 3**. Same as Table 1 but for SPOT 6.

The two BAPs have been identified in correspondence of elevated structures, some of them showing circular shapes, which display erosion and mass wasting on their sides. Both areas are exposed towards the lower Imhotep (Thomas et al., 2015) plain where the waste material is accumulated as boulders and debris (Filacchione et al., 2016a), which suggests that they originated from mass wasting. SPOT 6 is a cluster of some twenty small bright spots located in the Khepry region at the base of a scarp bordering a roundish flat terrace (Barucci et al., 2016). Also in this case it is very probable that the origin of the icy region is due to mass wasting from the nearby flat terrace.

The locations of the selected patches on the comet surface are indicated in Figure 1.

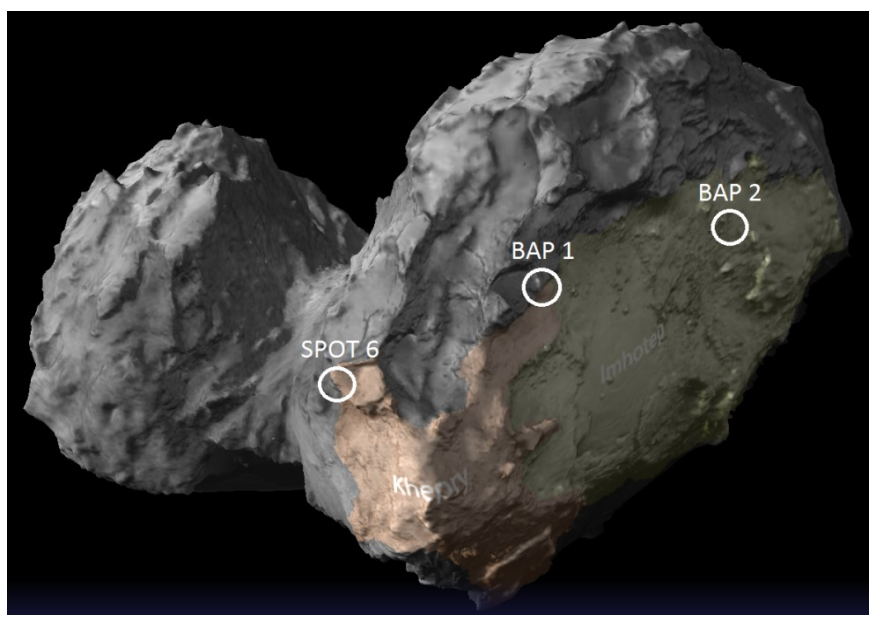

**Figure 1**. The three icy regions taken into account in this work are marked on the shape model of the comet surface (Preusker et al., 2015). Imhotep and Khepry regions are indicated and colored on the shape model.

**Analysis of spectral parameters**

Spectral analysis of VIRTIS data has shown that spectra across the surface present a positive ("red") VIS-IR slope and are basically featureless, except for the presence of a broad absorption feature around 2.9-3.6 µm. This absorption band is ubiquitous on the entire comet surface, and it is compatible with carbon-bearing compounds (Capaccioni et al., 2015). At large scale, most of the surface is depleted of water ice except for some specific regions where the presence of water ice is revealed by the shift of the absorption band toward shorter wavelengths (the band center moves from 3.2 to 3.0 µm), a slight increase in the band depth, a change of slope (the spectra are flatter, or "bluer") and a slight increase in albedo, which makes these spectra compatible with the presence of few percent of fine-grained water ice (De Sanctis et al., 2015; Filacchione et al., 2016b; Ciarniello et al., submitted).

By using both VIRTIS and OSIRIS data, Filacchione et al. (2016) and Barucci et al. (2016) observed a number of wider and more complex icy patches. For these icy regions, all the diagnostic absorption bands of water ice located at 1.05, 1.25, 1.5, 2.0, 3.0 µm (see Figure 2) are observed, and all the spectra display bluer levels of slopes in the VIS and IR ranges from the ice-free areas.

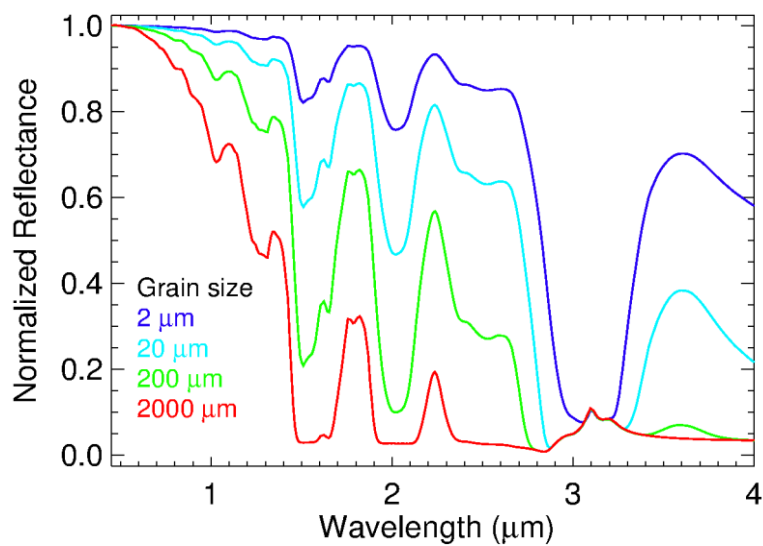

**Figure 2**. Spectra of pure water ice simulated with Hapke model (see Appendix) by means of optical constants provided by Warren et al. (1984), Mastrapa et al. (2008), Mastrapa et al. (2009), Clark et al. (2012), for different grain sizes. The diagnostic absorption bands of water ice located at 1.05, 1.25, 1.5, 2.0, 3.0 µm are clearly visible.

In order to derive quantitative information on the amount of ice, we have identified a number of spectral indicators that allow minimizing the spectral dimensionality, still preserving the relevant information contained in the spectra. The spectral indicators are the slopes in VIS and IR, and band areas.

The slope in the visible and in the infrared part of the spectra are calculated with a linear fit to the spectrum in the range 0.55 – 0.75 µm for the visible channel, and 1.0 – 2.35 µm for the infrared channel as shown in Figure 3.

The absorption band at 1.05 and 1.5 µm are not taken into account in this work because they are prone to artifacts, being the former located close to the junction between the two spectral channels, and the latter being affected by the junctions of the instrumental order sorting filters. The band areas of the absorption bands at 1.25, 2.0 and 3.2 µm (see Figure 4) are calculated as:

$$\int_a^b 1 - \frac{\text{reflectance}(\lambda)}{\text{continuum}(\lambda)} \, d\lambda$$

Being *a* and *b* the edges of the band: 1.16 - 1.38 µm, 1.83 - 2.24 µm, 2.62 - 3.60 µm, respectively. The continuum used to characterize the bands is calculated with a linear fit between *a* and *b*. The reflectance level correspondent to the band edges is calculated with a median of 7 spectels.

In those pixels where water ice is scarce or absent, the lowest value of the band area can be negative because the shape of the spectrum is concave while the continuum defined by the linear fit is a straight line. This is mainly the case for the band area at 1.25 µm. This does not affect our spectral analysis.

In the region 2.9-3.6 µm there is a superposition of the ubiquitous organics band and the water ice band. Although a variation of the band area at 3.0 µm can be due to either compound, we ruled out the contribution of the organics, as the increase in band area was always associated with a shift of the band center towards shorter wavelengths which clearly indicates an increasing content of water ice rather than organic material as already observed by Filacchione et al. (2016b) and Ciarniello et al. (submitted).

We mapped the spectral indicators across the selected areas, and in Figure 5 we show the distribution of the three band areas and the spectral slopes for a single acquisition of BAP 1. The spectral indicators are distributed according to different and well-defined patterns. The pattern shown by the band area at 2.0 µm is intermediate between the pattern shown by the band areas at 1.25 µm and 3.0 µm. The 1.25 µm band traces the areas with the largest ice abundance or largest grain size, while the 3.0 µm band is more widespread. This will be described in more detail in the next section.

The scatter plots in Figure 6 indicate the consistency of the results obtained for the various spectral indicators; the increase in the 2.0 µm band area is associated to a flatter IR spectrum pointing to an increase of the ice abundance, the same is true for the VIS and IR spectral slopes which consistently decrease with increasing abundance of ice, as described in next section.

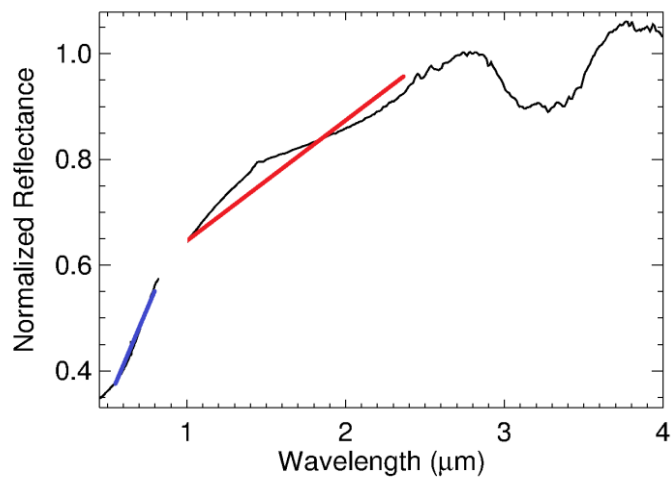

**Figure 3**. Reflectance spectrum of 67P normalized at 2.8 µm. This spectrum does not exhibit the typical water ice features. The spectrum in the range 0.5-1.0 µm is measured by the VIS channel, while the 1-5 µm range is covered by the IR channel. The broad absorption band at 3.2 µm is compatible with the presence of organic compounds. The slopes are calculated with a linear fit to the spectrum in the range 0.55 – 0.75 µm for the visible channel (blue line), and 1.0 – 2.35 µm for the infrared channel (red line).

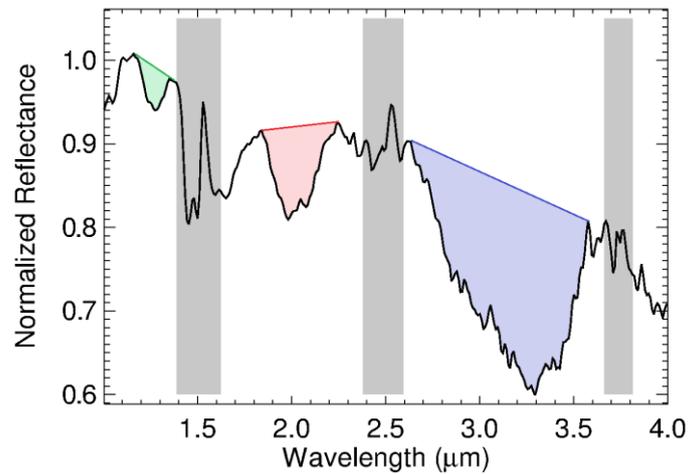

**Figure 4**. A reflectance spectrum normalized at 1.1 μm, showing significant water ice absorption bands, is depicted. The continuum in correspondence of the absorption bands at 1.25, 2.0 and 3.2 μm is calculated with a linear fit at fixed wavelengths. The band area under the continuum is highlighted. The gray boxes represent the spectral ranges affected by the junctions of the instrumental order sorting filters, which produce unreliable spectral features, and thus are not taken into account in the analysis.

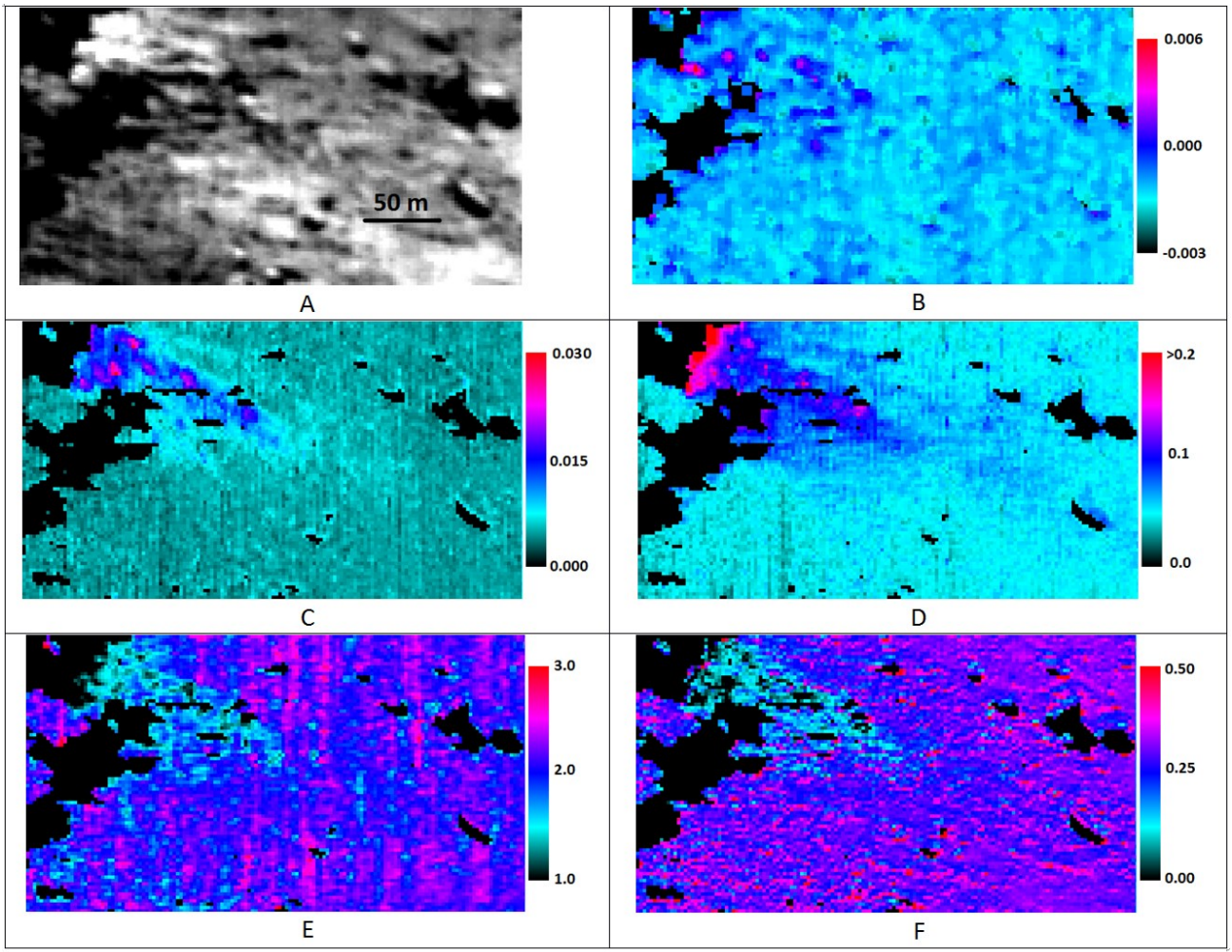

**Figure 5**. Analysis from acquisition 371998863 of BAP 1 (see Table 1). Panel A: context image (IR channel at 1.8 μm). Panels B, C, and D show the band areas (μm) of the absorption band at 1.25 μm, 2.0 μm and 3.0 μm, respectively. Panels E and F show the slopes (μm$^{-1}$) respectively in the VIS and IR channel. Non-illuminated pixels are marked in black and excluded from the analysis.

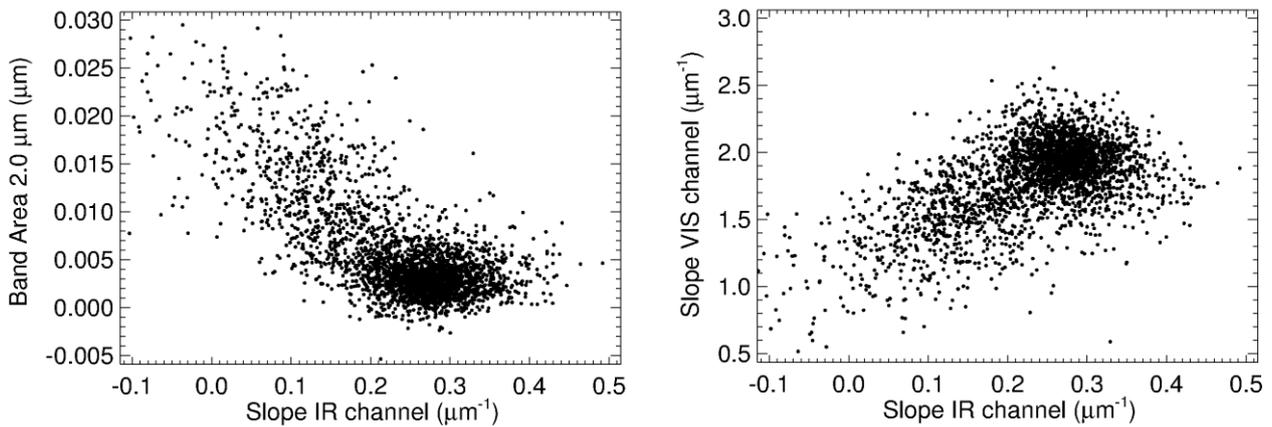

**Figure 6**. Scatter plot of the all pixels in acquisition 371998863, excluding those in shadow. Left and right panel show the good correlation between the slope in the IR channel with respectively the band area at 2.0 µm and the slope in the VIS channel.

**Spectral modeling**
To quantify the ice abundance, grain size and mixing modalities we have applied Hapke's radiative transfer model (Hapke, 2012) as described by Ciarniello et al. (2011). A similar approach was adopted in the two previous papers by Filacchione et al. (2016) and De Sanctis et al. (2015).
The model allows deriving the spectrophotometric properties of mixtures of several end-members characterized by their abundances and grain sizes. As the objective of this work is the study of water ice, we minimized the number of end-members by using only crystalline water ice and a "dark terrain" unit corresponding to the average spectrum of the comet's surface after the application of photometric correction (Ciarniello et al., 2015). Crystalline ice has been simulated using optical constants measured at T=160 K between 0.5 and 4 µm (Warren et al., 1984; Mastrapa et al., 2008; Mastrapa et al., 2009; Clark et al., 2012).
In addition we adopted two mixing modalities: a linear (*areal*) mixing in which the surface is modeled as a collection of patches of either pure water or pure dark terrain, and each photon interacts only with a single species. An *intimate* mixing in which single grains of the end-members are in contact with each other and the single photons can directly interact with both species.
The abundance and grain size of the water ice are free parameters. The abundance retrieved by the model has to be intended as the fraction of total cross section of the ice grains over the area subtended by the pixel.
Quantitative estimations on the deeper layers of ice cannot be directly retrieved because the reflected light measured by the instrument interacts only with the first layer of ice grains.
Abundance and grain size were retrieved using an iterative best-fit procedure to minimize the deviation between the model results and the measured spectra (see Appendix for further details on the model and the best-fitting procedure).

The results, in agreement with our previous work (Filacchione et al., 2016) are that the observed spectra require both areal and intimate mixture and, most interestingly, they require a bimodal distribution of icy grains: mm-sized grains in areal mixtures, and ~50 µm sized grains in intimate mixtures.
The left panel of Figure 7 shows an example of a fitted spectrum rich in water ice, which requires 0.9% water ice with 1800 µm grain size in areal mixture, and 2.2% water ice with 45 µm grain size in intimate mixture with the dark material. To illustrate the effect of the mixing modalities on the right panel we show the spectra produced by the two different populations. The most striking differences are the absence of the 1.25 µm band in the small grain size mixture (blue curve) and the saturation of all ice bands in the large grain spectrum (red curve).

In order to characterize the influence of ice abundance and grain sizes of both grain populations on the overall shape of the spectra we performed specific simulations by taking into account the typical retrieved values, as shown in Figure 8. The simulated spectra are normalized in order to highlight the variations of the spectral indicators as a function of the parameters of the model:
- The slope in the VIS channel is only affected by variation in abundance for both populations of grains;
- The slope in the IR channel is affected by all four parameters;
- The band area at 1.25 µm is only affected by properties of the ice with large grains in areal mixture;
- The band area at 3.0 µm is only affected by properties of the ice with small grains in intimate mixture;
- Band area at 1.5 µm (not taken into account in this work) and band area at 2.0 µm are affected by all parameters.

While it's relatively common to find ice patch spectra without the 1.25 µm absorption band (only intimately mixed water ice), spectra which clearly present a saturation of the icy features (see red curve of Figure 7) are very uncommon or absent among the pixels constituting the analyzed areas. This means that the patches of mm-sized water ice are usually embedded in a surface of intimately mixed and finer grained ice. This is indeed what is observed in Figure 5 (panels B and D).
In Figure 9 two different collections of measured spectra show the effect of a variation of the abundance of both populations of water ice. Model used to retrieve the abundances is the same used to produce the simulated spectra in Figure 8.

The variety of the measured spectra can be mainly attributed to a variation of the abundance and partially to the grain size of the water ice within the spots. This is shown in the scatter plots in Figure 10, which reports the case of the acquisition 371998863 (the same as for Figures 5 and 6) as an example: the abundance of water ice in areal mixture (panel A of Figure 10) is in good correlation with the band area at 1.25 µm. By grouping the symbols in the Figure according to their grain size it seems that for the pixels with grain size < 2000 µm the upper limit is ~1% of abundance, while it is higher for pixels with larger grain size. Panels B, C, D of Figure 10 show the correlation between the abundance of the water ice in intimate mixture with the band area at 2.0 µm, the band area at 3.0 µm, and the slope in the IR channel, respectively. For the 2.0 µm band area and the IR slope case, the different grain sizes lead to different lines of correlation. For the 3.0 µm band area, the different groups seem to follow the same line, but they are limited by different upper bounds. These three parameters are marginally affected by the variation of properties of the mm-sized grain population because of the lower amount of the latter. For the purpose of this analysis only spectra with a band area at 2.0 µm larger than 0.005 µm are taken into account. This threshold is equal to the mean level + 3 standard deviations of the band area, as calculated in a region outside of the icy patches. Possible interpretations of the results obtained here are discussed in the next sections under the general context of the temporal evolution of the icy patches.

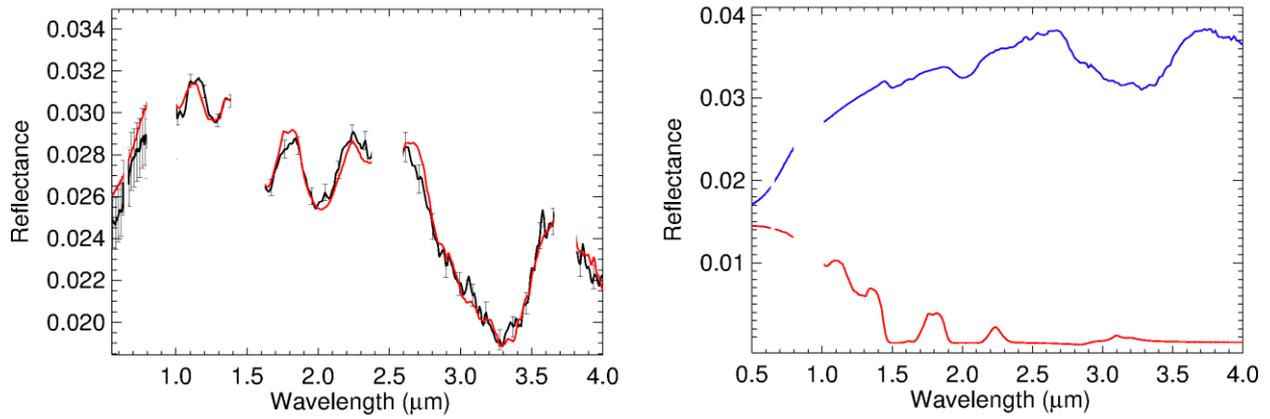

**Figure 7**. Left panel: VIRTIS reflectance spectrum (sample 25, line 20, spacecraft clock time 371998843) in black. Error bars are calculated as described by Raponi (2014). Best fit in red: 0.9% water ice with 1800 μm grain size in areal mixture, and 2.2% water ice with 45 μm grain size in intimate mixture with the dark material. Spectral ranges affected by the junctions of the filters are not taken into account in the analysis. Discrepancy between best fit and the VIS channel can be due to the not perfect co-registration between VIS and IR channels. In the right panel the two populations of the water ice retrieved with the model have been decoupled: in red a scaled spectrum of pure water ice with 1800 μm grain size, in blue an intimate mixture of 2.2% water ice with 45 μm grain size and 97.8% dark terrain.

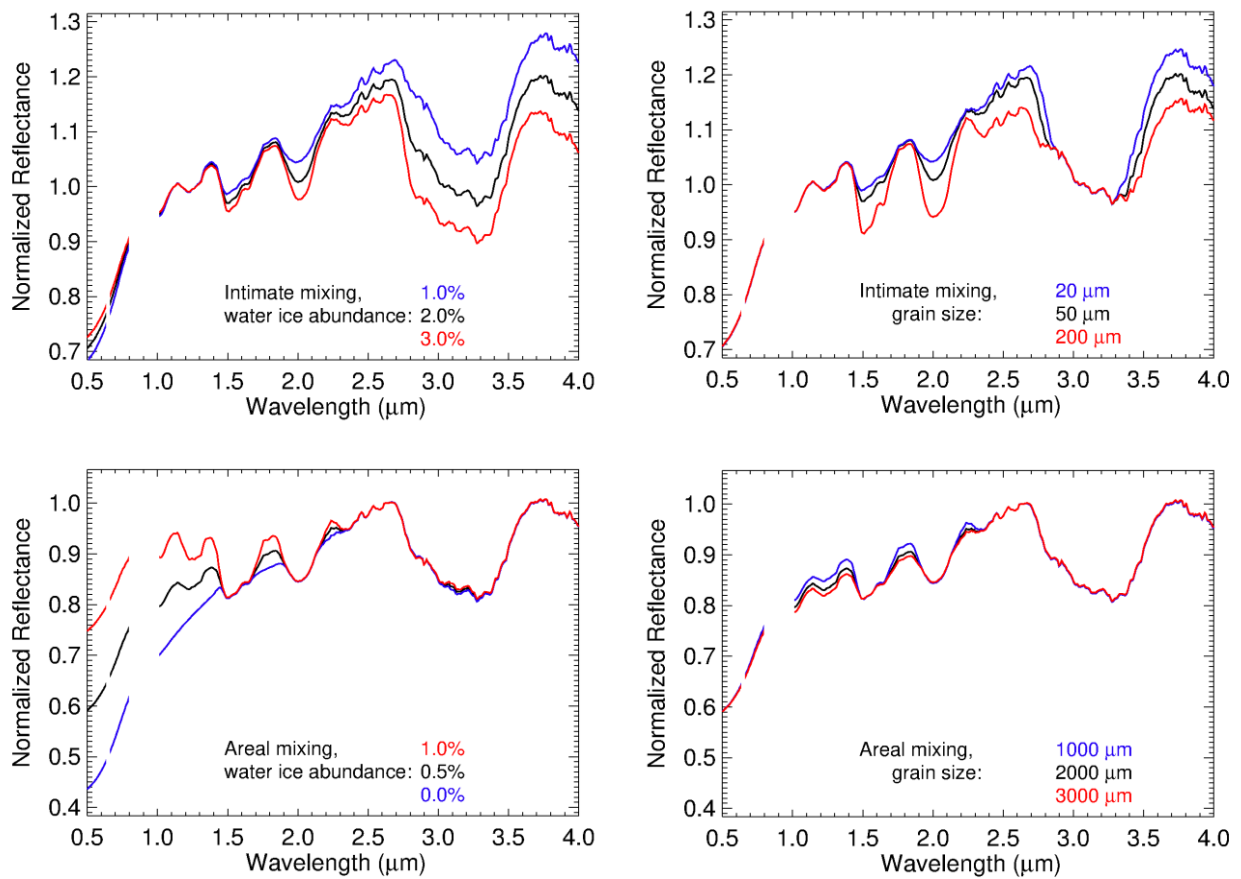

**Figure 8**. Simulated reflectance spectra are shown to highlight the effect of variation in abundance and grain size of the water ice. The black curve represents an intimate mixture of dark terrain and 2% of water ice with 50 μm grain size, which in turn is in areal mixing with 0.5% water ice with 2000 μm grain size. The red and blue curves are simulated by varying one of these parameters at a time, as indicated: the panels on the left show the effect of a variation in abundance, and the panels on the right a variation in grain size, for both populations of water ice: intimate mixing in the upper panels (spectra normalized at 1.16 μm), and areal mixing in the lower panels (spectra normalized at 2.62 μm). The spectral ranges are not taken into account in correspondence of artifacts that have prevented the definition of the albedo of the dark terrain.

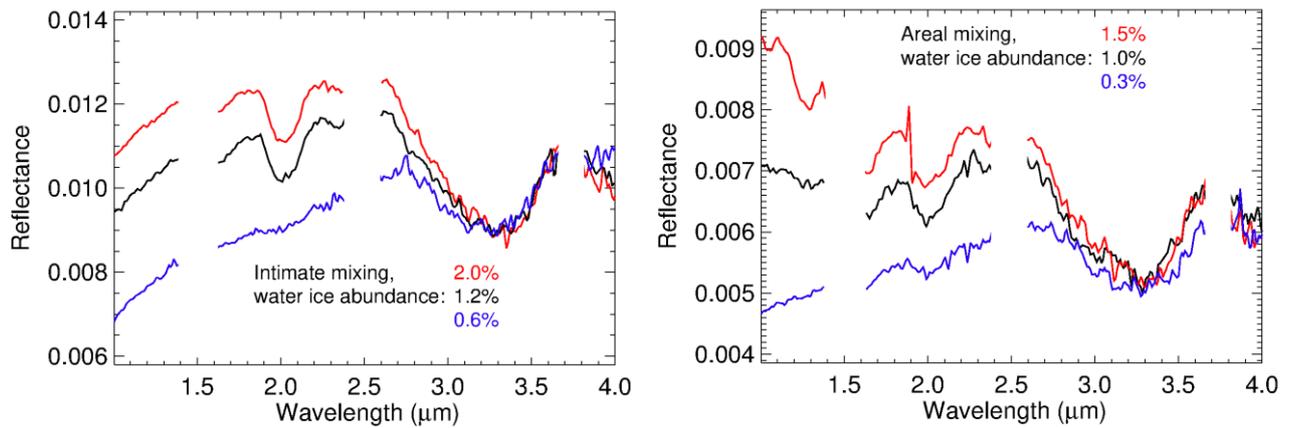

**Figure 9**. According to the model, the collected measured spectra show the effect of a variation of the abundance of water ice in intimate mixture (left), and the effect of variation of the abundance for the areal mixtures (right). The retrieved abundance is shown in the panels.

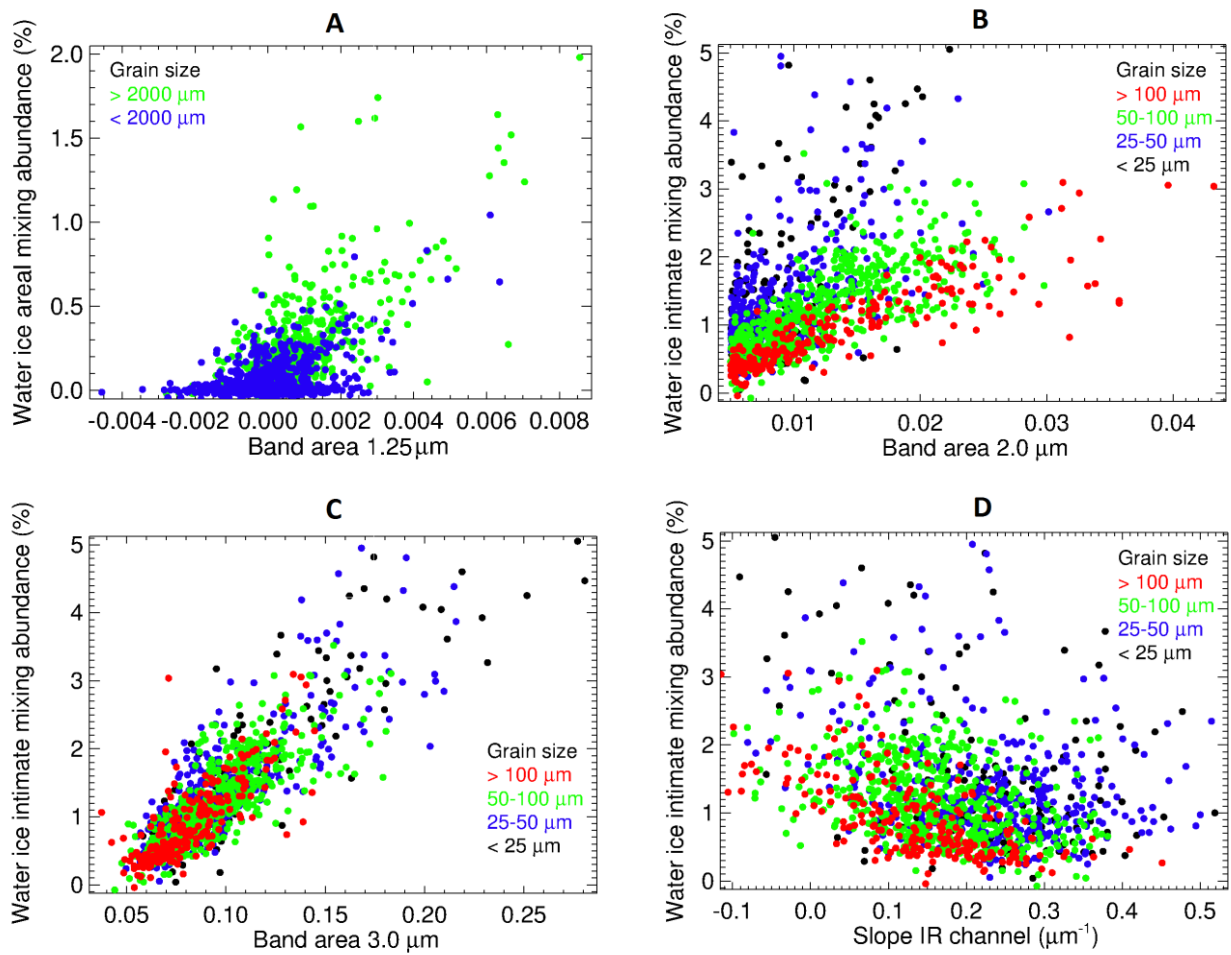

**Figure 10**. Analysis of acquisition 371998843. Spectra selected are those with band area at 2.0 μm larger than 0.005. Panel A: Abundance of water ice in areal mixture as a function of the band area at 1.25 μm. Panels B, C, D show the correlation between the abundance of water ice in intimate mixture with respectively the band area at 2.0 μm, the band area at 3.0 μm, and the slope in the IR channel. The symbols are marked with different colors according to their grain size as indicated in the panels.

**Water ice property retrieval by means of spectra in the visible range**

The aim of this section is to give a useful guideline for the retrieval of the icy patches' properties by means of instruments operating only in the visible spectral range.

Thanks to the Hapke model we can retrieve the water ice abundance by modeling the normalized spectral slope. We do not take into account the absolute signal level because it can be affected by uncertainties on the radiometric and photometric accuracy as well as errors on the local geometry information, due to unresolved shadows and roughness.

As shown in Figure 8 the slope in the VIS channel is sensitive to the water ice abundance and the type of mixing (areal or intimate), while it is not affected by the grain size. In principle we cannot distinguish between the two types of mixing by just the slope in the VIS channel, and thus we can only give lower and upper limits by assuming water ice in areal or intimate mixture with the dark material, respectively (black and red lines in Figure 11). However, from the analysis performed on the three icy patches discussed in the present work, we obtain as a general result that the ratio between the abundances of water ice in intimate and areal mixture is in the range 1 - 10 (see next sections and upper panels of Figures 15, 16, and 17). By assuming this range of ratios, we can reduce the gap between the lower and upper limit (blue and green line in Figure 11).

The phase function should also be accounted for by the model, as it can produce slope variations of the order of +0.007 $\mu m^{-1}$ / deg, according to Ciarniello et al. (2015).

Moreover, the dark terrain used as a spectral end-member could contain itself a small percentage of water ice, and thus the water ice abundance could be underestimated.

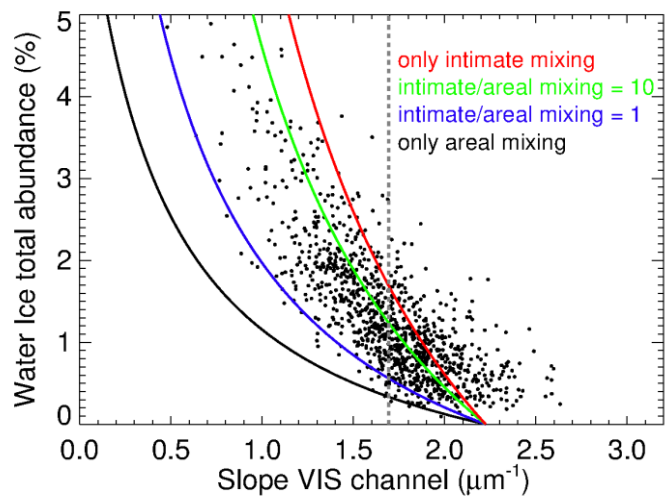

**Figure 11**. Curves: theoretical abundance of water ice as a function of the slope in the VIS channel, as calculated by means of the Hapke radiative transfer model at a phase of 95.0° (same phase as acquisition 371998843). The red and black lines correspond to water ice entirely in intimate or in areal mixture with the dark terrain, respectively. The blue and red lines correspond to the ratio between the amount of water ice in intimate and areal mixture equal to 1 and 10, respectively. Dots: Water ice total abundance as a function of the measured slope in the VIS channel, for the acquisition 371998843. Spectra selected are those with band area at 2.0 μm larger than 0.005. Abundance is retrieved as described in previous section by means of the spectra in the IR channel. Dashed gray line: average slopes of the dots. The width of the dashed line is equal to the error on the average slope.

We take into account the acquisition 371998843 in order to compare the theoretical curves shown in Figure 11 with the abundance retrieved as in previous section by means of the spectra in the IR channel (the points of Figure 11). The ratio between the water ice abundance in intimate and areal mixing is close to 10 for this acquisition (see upper panels of Figure 15). This is consistent with the dispersion of the points shown in Figure 11 around a curve close to the green one.

By taking the average of the VIS slope obtained for the points shown in Figure 11 we can constrain the amount of water ice of the patch in the range 0.6 – 1.2 %, assuming the ratio between the abundances of water ice in intimate and areal mixture in the range 1 - 10. The total abundance retrieved with spectra in IR range is 1.1 % (see upper left panel of Figure 15), thus demonstrating the consistency of the result obtained with spectra in the VIS channel with that obtained with spectra in the IR channel.

**Temporal evolution of ice-rich areas**

The collected acquisitions listed in Tables 1, 2, and 3 allow following the temporal evolution of the three icy regions analyzed in this work. According to the previous discussion and as shown in Fig. 4, we selected the absorption band at 2.0 µm as a reliable indicator of the water ice presence. Each hyperspectral cube was orthorectified and projected on a cylindrical map. The temporal evolution of the 2.0 µm band area in each ice-rich region, BAP 1, BAP 2 and SPOT 6, is shown in Figures 12, 13 and 14.

To give a quantitative estimate of the amount of ice present in the region under observation, the spectra with band area at 2.0 µm larger than 0.005 are taken into account. The retrieved water ice abundances are averaged, weighted by the area covered by each pixel, calculated as Area = resolution$^2$ / cos(emission angle). The result is the fraction of the total area of the patch in which the solar flux interacts with the icy grains, which is a proxy for the surface density of water ice (upper left panels of Figures 15, 16 and 17). This abundance is multiplied by the total area subtended by the pixels taken into account, to obtain the total cross section of the water ice (upper right panels of Figures 15, 16 and 17). The lower panels of Figures 15, 16, and 17 show the grain size retrieved for the two populations of grains as a function of time.

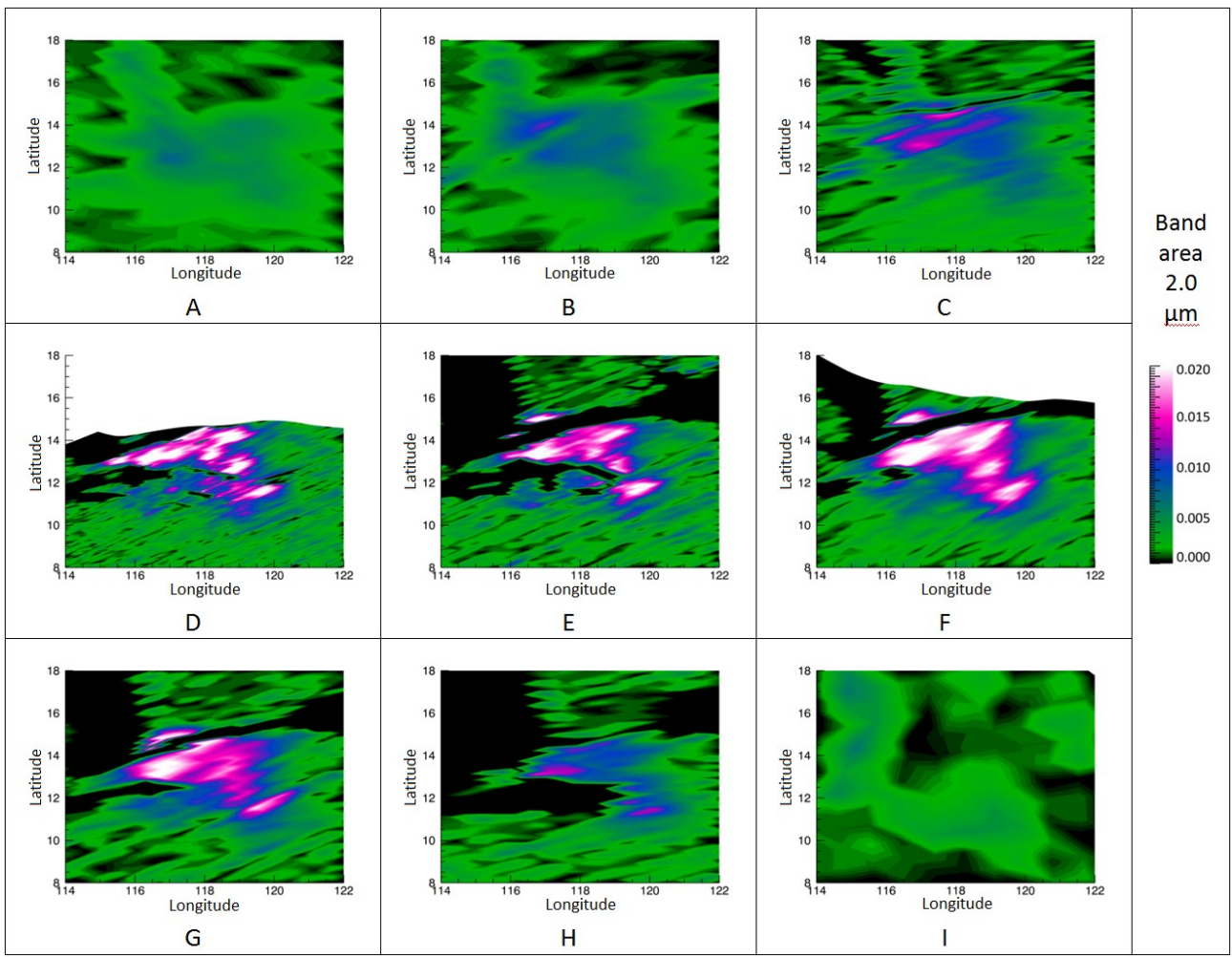

**Figure 12**. The band area at 2.0 µm is mapped for each acquisition of BAP 1. The panels represent the acquisition listed in Table 1. The increase of the extension and intensity of the feature in the first 6 panels, and then their decrease are quite evident. In the last panel the feature is very weak or absent. The dark areas are in shadow. The white areas outside the edge of the map in panels D and F are outside the field of view of the acquisitions.

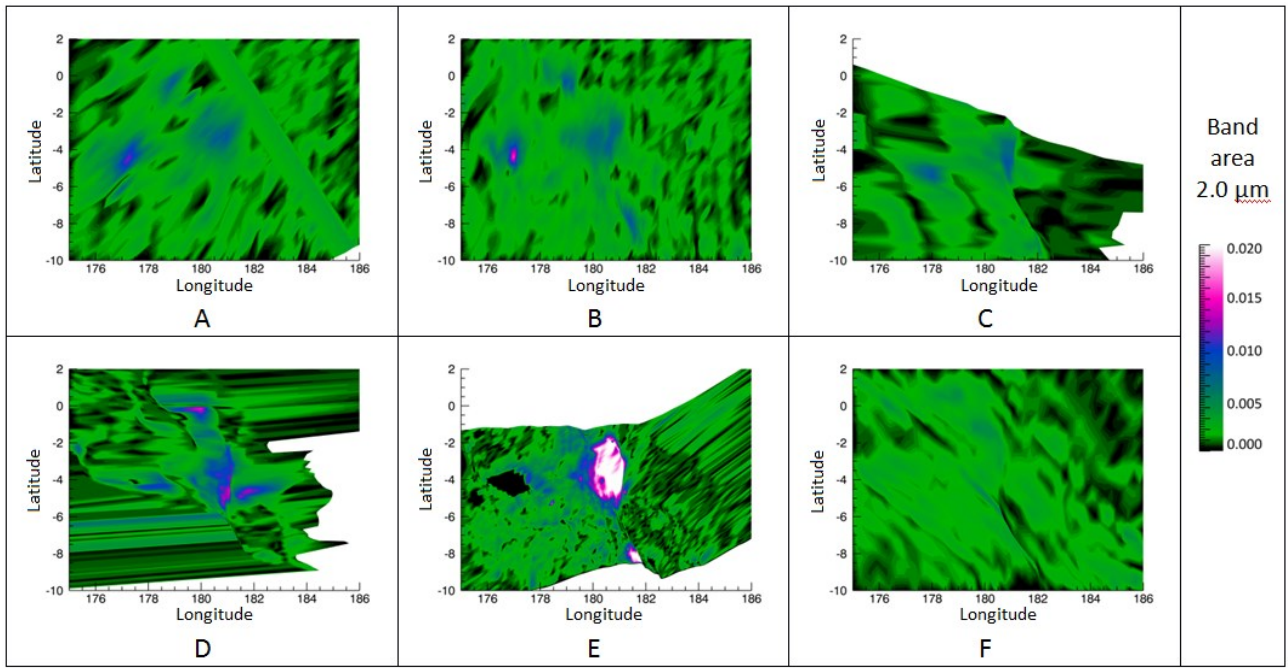

**Figure 13**. Same as Figure 12 but for BAP 2.

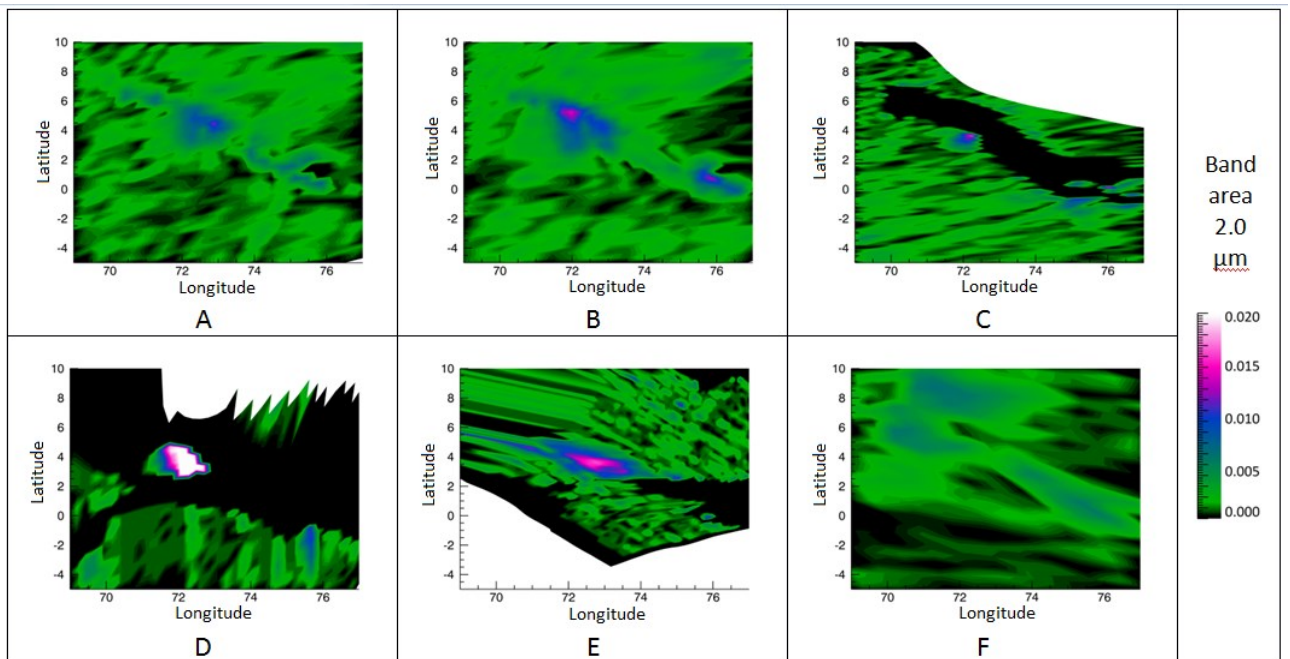

**Figure 14.** Same as Figure 12 but for SPOT 6.

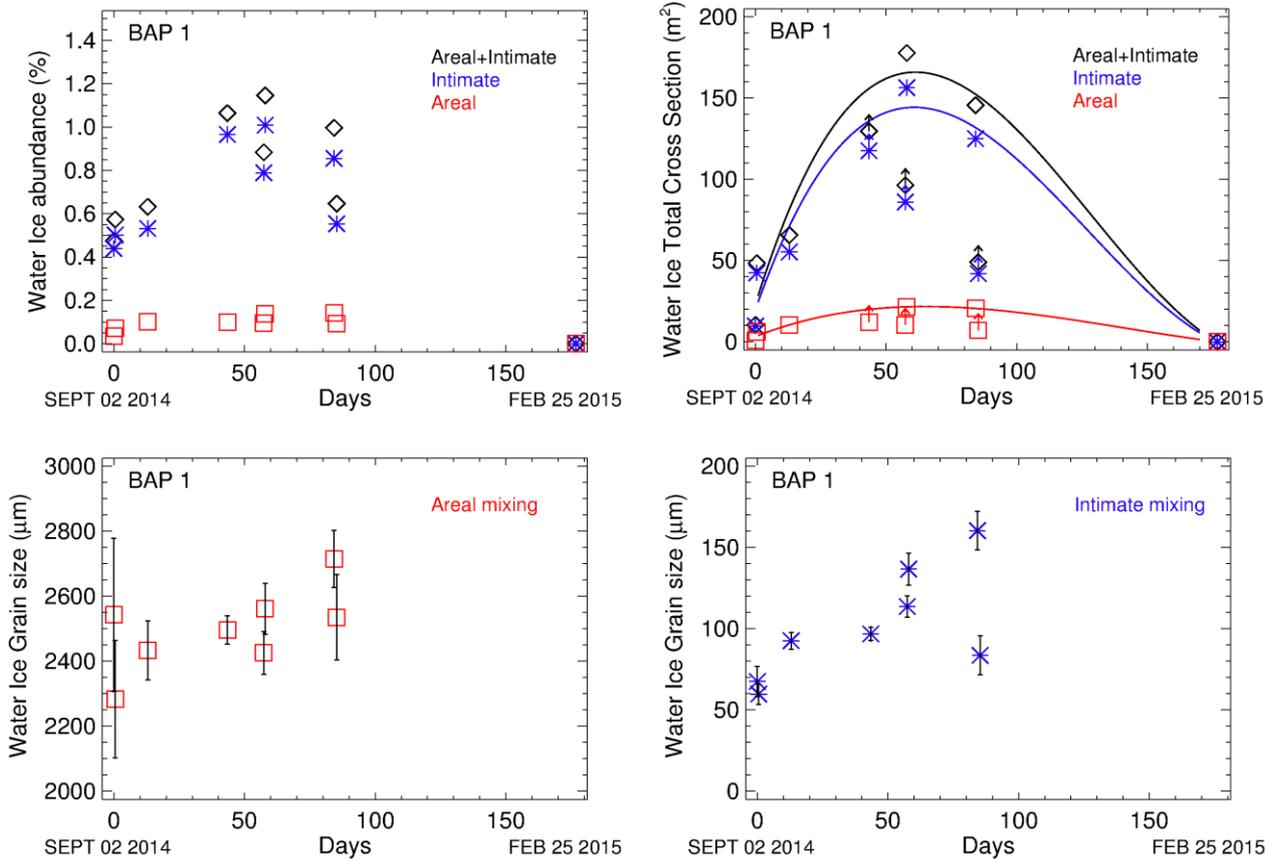

**Figure 15**. BAP 1: water ice abundance (upper left panel) and total cross section of water ice (upper right panel) is plotted as a function of time for the two types of mixtures modeled. The latter could be affected by the limited view of the icy region because of shadows, or because part of the region is outside the field of view of the instrument: those symbols are marked with an arrow indicating a lower limit. The lines are produced with a polynomial fit of the points (without including those with the arrow). The bell shaped curve is evident for both populations of water ice. The total amount of the water ice in areal mixture is always a small portion of the total amount of water ice which covers the surface. The errors on the retrieved values (not shown for the sake of clarity) are of the order of the size of the symbols.

Lower panel: the mean grain size is plotted as a function of time.

The errors are calculated as the standard deviation of the values divided by the square root of the number of selected pixels.

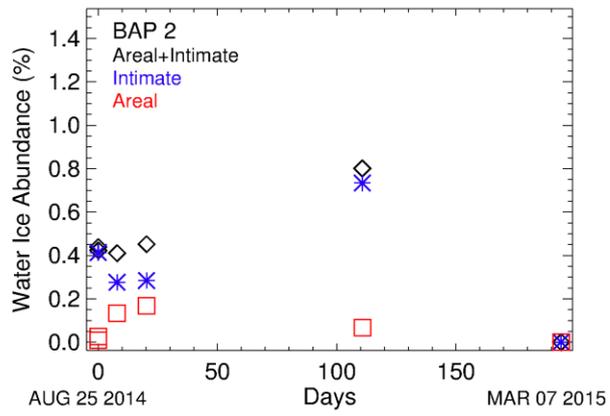
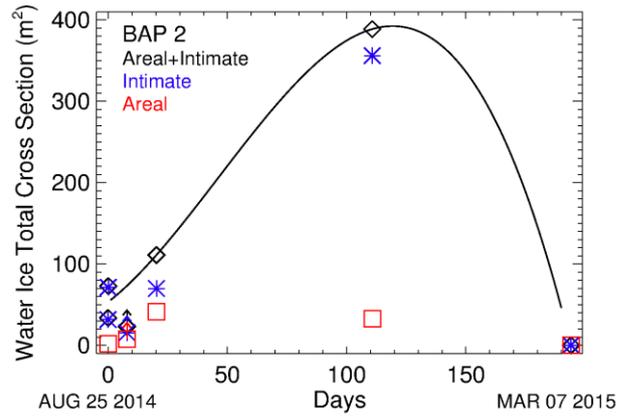
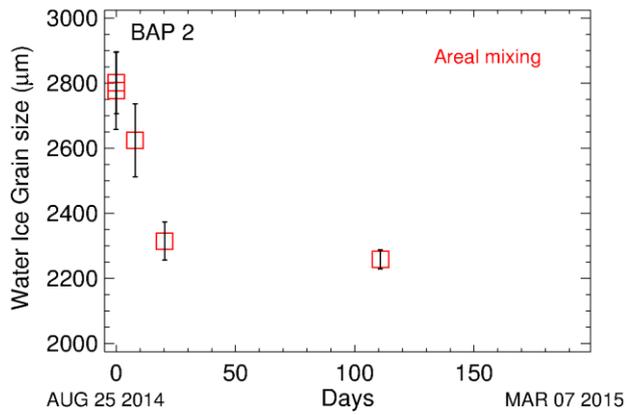
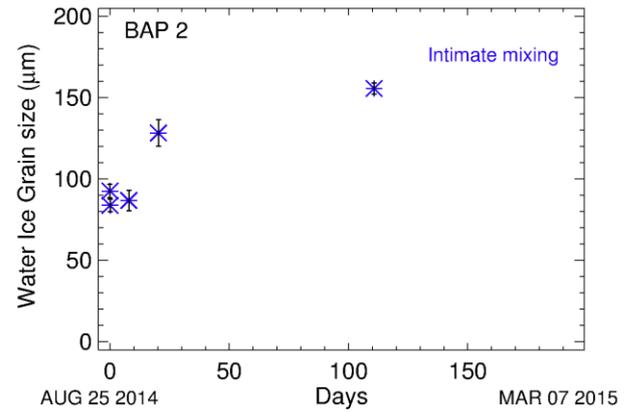

**Figure 16**. Same as Figure 15 but for BAP 2.

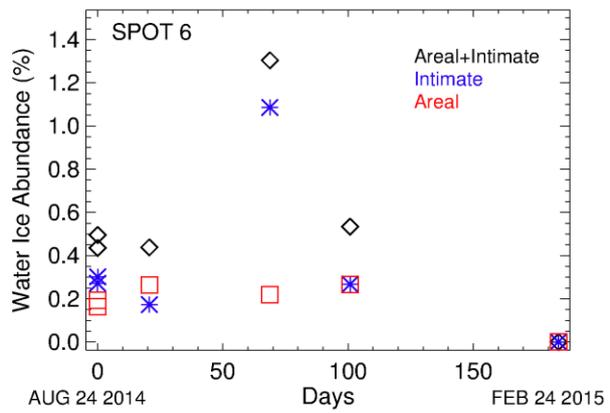
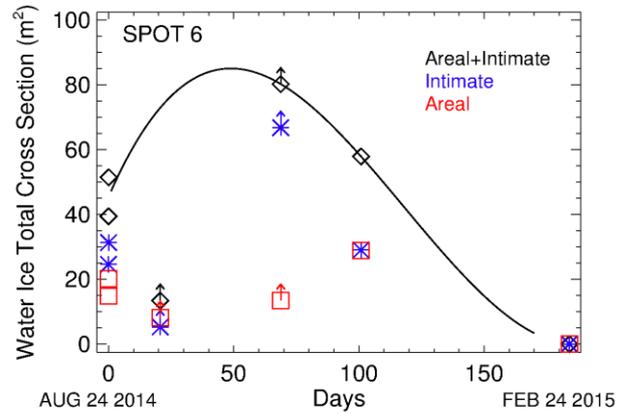
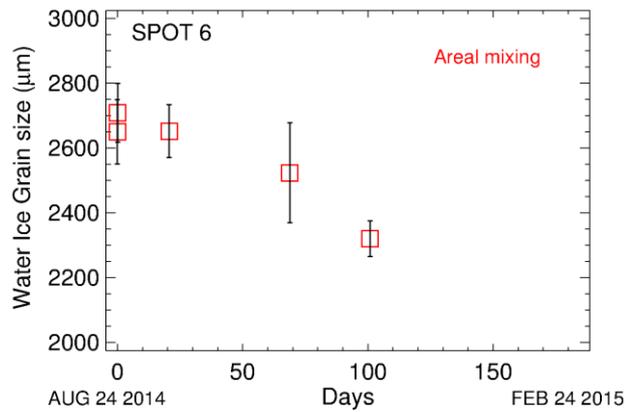
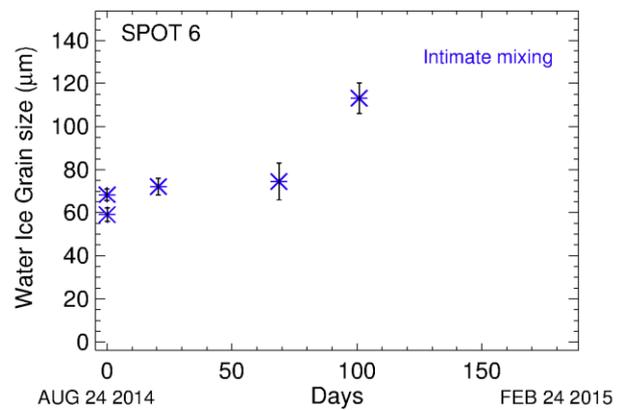

**Figure 17**. Same as Figure 15 but for SPOT 6.

**Discussion**

The most striking aspect of the observations presented in this work is that all three ice-rich regions have a life cycle in which the icy patch increases in size and abundance, reaches a maximum and is followed by a rapid decrease after which, eventually, the ice disappears from the surface. The three areas analyzed are all very close to the equator, which is consistent with their similar thermal history and evolution.

The determined lifecycle allows us to realistically exclude a diurnal cycle as the major driver for the observed variability and, indeed as reported in Tables 1, 2 and 3, the observations cover a restricted range of local times and incidence angles.

However, significant variations on smaller time scales can be appreciated both in the extension and intensity of the icy features. As an example panels G and H of Figure 12 concern two acquisitions of BAP 1 which are close in time (24 hours, two comet's rotations) but the latter presents a significantly smaller extension and lower intensity of the 2.0 μm band. This is also confirmed by the retrieved abundances (upper panels of Figure 15).

This difference cannot be accounted for entirely by the diverse shadowing. Although we cannot exclude uncertainties due to the different resolution and viewing geometry of the acquisitions, it is reasonable to explain such variation with a diurnal cycle, which produces short-term behavior, overlapped to the seasonal cycle. The spatial distribution of the 3.0 μm and 1.25 μm band area values could support the understanding of the temporal stability of the patch: as discussed above, the 3.0 μm band area (panel D of Figure 5) is an indicator of the presence of the finer intimately mixed icy grains. Their concentration seems to follow the shadow, being larger closer to the non-illuminated areas, indicating a sublimation of the finer icy grains when exposed to the solar light. Conversely, the 1.25 μm band area is an indicator of larger grains in areal mixing (panel B of Figure 5), which are grouped in small spots independently from the shadow position, indicating their temporal stability with respect to the solar input, at least in the diurnal time scale.

Most of the water ice is in the form of finer grains, in intimate mixture with the dark material. This population is the main driver of the variability of the amount of water ice (see upper panels of Figures 15, 16, 17). In contrast, ice in mm-sized grains appears very locally concentrated in specific spots embedded in the more widespread finer grains.

Except for BAP 1, the average size of the mm-sized grain population seems to decrease as a function of time (lower left panels of Figures 15, 16, 17), most probably because the sublimation produces erosion of the grains. Conversely, the average size of the ~50 μm sized grain population increases with time for all patches. This can be explained by the complete sublimation of the smaller grains, which causes the size distribution to shift towards larger sizes. This is also confirmed by their spatial distribution shown in panel C of Figure 10: the pixels with the larger amount of water ice (blue and black dots) are populated by the finer grains and vice versa, indicating that where the sublimation has been less effective the water ice is still abundant with a small average grain size. On the other hand, inside the icy spots with mm-sized grains, more affected by grain erosion than complete sublimation, the pixels present larger abundances and larger grain sizes (green dots of panel A in Figure 10) where the solar input has been less intense.

**Conclusion**

As reported by Filacchione et al. (2015) and Barucci et al. (2016) the observed patches occur in areas affected by mass wasting and erosion; consequently, the three ice-rich regions very likely represent the occasional exposure of sub-surface layers fairly rich in water ice. This is a common finding on 67P where ice is predominantly absent and only temporary present at the surface, when exposed due to occasional erosional events or in form of recondensed ice in specific location (De Sanctis et al, 2015).

In addition, in all these areas, along with ice grains of tens of microns intimately mixed with the dark component, we observed also areas with mm-sized grains, which make a relevant contribution to the overall water ice abundance. The presence of mm-sized grains is quite unexpected at the surface of the nucleus because, as reported by many authors (Yang et al., 2009; Sunshine et al., 2007; Protopapa et al., 2014; De Sanctis et al., 2015; Ciarniello et al., submitted), modeling of observed coma and nucleus surface spectra seems to indicate that the typical size range of the ice grains observed in comets never exceeds tens of micron.

Filacchione et al. (2015) already identified mm-sized ice grains in BAP 1 and BAP 2. The addition of SPOT 6 from a completely different region indicates that the presence of larger grains in subsurface layers is a common feature of the water ice's physical status. The large size possibly derives from physical processes acting within deeper layers of the nucleus, as for instance by the growth of secondary ice crystals from vapor diffusion in ice-rich colder layers, and/or by ice grain sintering (see Filacchione et al., 2016a, and references therein).

The exposure of deeper layers is consistent with their occurrence in "active" areas where falls or landslides could have caused the occasional exposure of water ice rich layers.

After the initial exposure of the ice, the activity of the affected area increases thus causing dust removal powered by sublimation, which provides a positive feedback on the exposure itself. The process develops as the solar flux increases, and it reaches a turning point when the exposure rate is overcome by the sublimation rate, until the complete sublimation of the patch.

The similarity in the lifecycles of the three patches analyzed in this work (180 days) and the similarity in the ice grains size distribution, seems to indicate that water ice is rather evenly distributed in the subsurface layers and mixed to the ubiquitous dark terrain. Furthermore, our present findings also indicate the absence of large water ice reservoirs in the subsurface. This is confirmed by the recent work of Filacchione et al, 2016b and Ciarniello et al., 2016 where an overall and diffuse increase in the abundance in water ice has been derived from the VIRTIS observations spanning the period August 2014 through May 2015. Indeed, the consistent erosion rate of the nucleus of 67P (Bertaux, 2015; Keller et al., 2015), would allow exposure of large-scale dis-homogeneities, and in particular of regions highly enriched in water ice, which would last for a longer timescale, if present.

The knowledge of the temporal evolution of the ice-rich regions identified, can constrain the thickness of the layer and the amount of water released to the coma during the sublimation process: this will be the subject of a following work.

**Appendix**

*Hapke model and fitting procedure description*
A quantitative spectral analysis of the composition has been performed using Hapke's radiative transfer model (Hapke, 2012) as described by Ciarniello et al. (2011).

Eq. 1A $$r(i,e,g) = \frac{w(\lambda)}{4\pi} \frac{\mu_0}{\mu_0 + \mu} \left[ B_{SH}\, p(g,\lambda) + H(w,\mu_0)H(w,\mu) - 1 \right] S(i,e,g,\theta) B_{CB}$$

Where $r$ is the bidirectional reflectance, *i, e, g* are the incidence, emission and phase angle, respectively; *w* is the single scattering albedo (SSA); *p(g,λ)* is the single particle phase function; $\mu_0$, $\mu$ are the cosines of the incidence and emission angles; *H(w,μ)* the Ambartsumian-Chandrasekhar functions describing the multiple scattering components; $B_{SH}$ the shadow hiding opposition effect term; $B_{CB}$ the coherent back-scattering opposition effect; $S(i, e, g, \theta)$ the shadow function modeling large scale roughness and $\theta$ the average surface slope.

During the VIRTIS-M observations described in this paper the solar phase was always > 35°. As a consequence, the shadow hiding effect is negligible. This implies posing the terms $B_{SH}$ and $B_{CB}$ equal to 1. The other photometric parameters in the equation are fixed as resulting from Ciarniello et al. (2015).

The geometry information is derived from the information on the digital shape model of the comet (Preusker et al., 2015), the spacecraft attitude, the comet's attitude, and the relative positions.

The single scattering albedo *w* can be modelled as a mixing between the dark terrain of the comet and water ice. The former was defined by Ciarniello et al. (2015), for the latter the SSA is calculated from the optical constants (Warren et al., 1984; Mastrapa et al., 2008; Mastrapa et al., 2009; Clark et al., 2012) and from the grain size by means of the Hapke compositional model (2012).

We have used areal and intimate mixing modalities in the simulations: in the areal mixing the surface is modelled as patches of pure water ice and dark terrain, with each photon scattered within one patch. In this case the resulting reflectance is a linear combination of the reflectance of the different end-members.

In the intimate mixing model the particles of the two end-member materials are in contact with each other and both are involved in the scattering of a single photon. In intimate mixing the single scattering albedo of the mixture is the weighted average, through their relative abundance P, of the Dark Terrain (DT) and Water Ice (WI) single scattering albedos.

Most of the spectra inside the patches require both kinds of mixtures and two different grain sizes of water ice. Thus the resulting reflectance $r(\lambda)$ is given by:

Eq. 2A     $r(\lambda) = r_{WI\_A}(\lambda) \, P_{WI\_A} + r_{WI\_DT\_I}(\lambda) \, (P_{WI\_I} + P_{DT})$,
    with: $r_{WI\_DT\_I} = r_{WI\_DT\_I} \, (w(\lambda)_{DT} \, P_{DT} + w(\lambda)_{WI\_I} \, P_{WI\_I})$,

Where *r* is the bidirectional reflectance (Eq. 1A) and the suffix "A" and "I" state for the areal and intimate mixture.

The best-fitting result is obtained by applying the Levenberg-Marquardt method for nonlinear least squares multiple regression (Marquardt, 1963)

Free parameters of the model are:
- Abundance and grain size of the water ice for both intimate and areal mixtures.
- A multiplicative constant of the absolute level of reflectance of the model, in order to account for uncertainties on the radiometric and photometric accuracy as well as errors on the local geometry information, due to unresolved shadows and roughness.
- A slope added to the model in order to better fit the measured spectrum in the IR channel: in some cases, the measured spectra present an artificial slope where high signal contrast is measured between adjacent pixels, like regions near shadows. This is due to the increasing full with half maximum of the spatial point spread function toward longer wavelengths (Filacchione, 2006).
- Temperature and effective emissivity (Davidsson et al., 2009). The latter is the product of the directional emissivity (see next section) and a free parameter used to account for unresolved shadow and the structure of the surface (Davidsson et al., 2009). Its interpretation is outside the scope of the present work.

The total radiance is modeled by accounting for both the contributions of the reflected sunlight, and the thermal emission:

Eq. 3A        $Rad(\lambda) = r(\lambda) * F_{\odot} / D^2 + \varepsilon_{eff} * B(\lambda, T)$

Where $r(\lambda)$ is the Hapke bidirectional reflectance (Eq. 1A), $F_\odot$ is the solar irradiance at 1 AU, $D$ is the heliocentric distance (in AU), $\varepsilon_{eff}$ is the effective emissivity, $B(\lambda, T)$ is the Planck function.

**Emissivity**

In order to model the thermal emission we take into account the directional emissivity ($\varepsilon_\lambda$) relation (Hapke, 2012) which is a function of the single scattering albedo ($w$):

Eq 4A $\qquad \varepsilon_\lambda = H(w, \mu) * \gamma(w)$

being $\mu$ the emission angle cosine, and $H$ the Ambartsumian-Chandrasekhar function. According to Kirchhoff's law of thermal radiation, the emissivity and reflectance spectral contrasts (difference between continuum and band signal) have opposite signs. The higher is the thermal emission in the spectral range of the absorption band, the lower is the band depth (by summing the reflectance and the thermal emission), and it can even become an emission feature. By means of Eqs. 1A-4A, the spectral contrast has been simulated for the sum of the reflected and thermally emitted fluxes. In Figure 1A two different conditions are simulated on the base of the viewing geometry of the icy patches analyzed in this work and the temperatures obtained by Filacchione et al. (2015) and Barucci et al. (2016). As a result the emissivity spectral contrast can be considered negligible with respect to the reflectance spectral contrast.

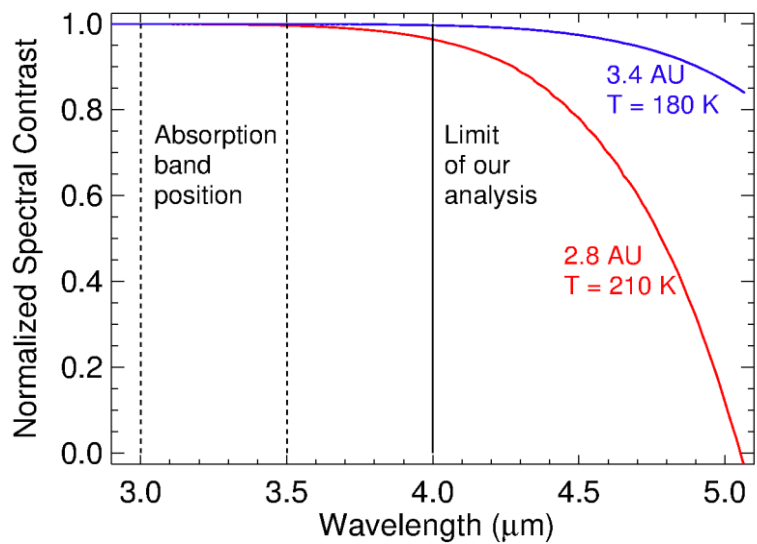

**Figure 1A**. The plots show the sum of the simulated spectral contrasts of reflected and emitted light. It is normalized to the value at shorter wavelengths (< 3.0 µm) for which thermal emission is not observed. Two conservative thermal conditions are simulated. For T = 180 K (blue line) the effect of the thermal emission is negligible in the spectral range taken into account in this work. For T = 210 K (red line) a possible absorption band at 5.0 µm would be flattened, and longward of that it would become an emission feature. Shortward of 4.0 µm, which is the limit of our analysis, the spectral signatures are affected as much as 5%. For the absorption band measured at 3.0 - 3.5 µm the effect is always negligible.

**Uniqueness of the solutions**

In order to produce the best fit we should evaluate if the solutions steadily converge, or if there are secondary solutions. This evaluation is made by the investigation of the $\chi^2$ function defined in the N-dimensional space, where N is the number of the free parameters. As an example we report in Figure 2A the projections of this space around the best fit of Figure 5: the icy spectrum for which the resulting water ice in areal mixture is 0.9 % and 1800 µm grain size, and 2.2 % at 45 µm grain size of water ice in intimate mixture with the dark terrain. As a result of these investigations significant secondary minima have never been found. Thus we can rely on the uniqueness of the solution.

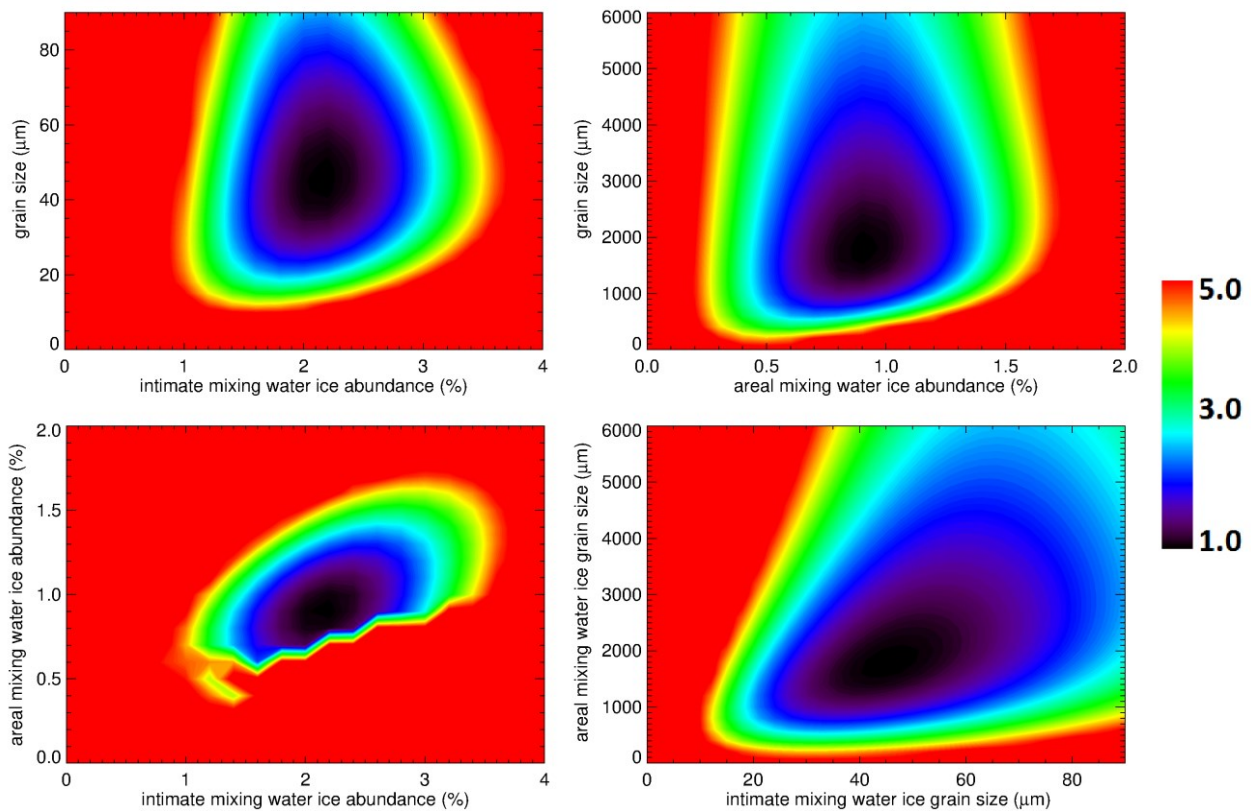

**Figure 2A**. In the four panels the $\chi^2$ function is plotted as a function of water ice abundances and grain sizes of the two populations of water ice. The absence of secondary minima confirms that the solutions are stable and there are no other sets of parameters which can produce models as good as the one retrieved.

**Acknowledgements**
We thank the following institutions and agencies for support of this work: Italian Space Agency (ASI, Italy) contract number I/024/12/1, Centre National d'Etudes Spatiales (CNES, France), DLR (Germany), NASA (USA) Rosetta Program, and Science and Technology Facilities Council (UK). VIRTIS was built by a consortium, which includes Italy, France, and Germany, under the scientific responsibility of the Istituto di Astrofisica e Planetologia Spaziali of INAF, Italy, which also guides the scientific operations. The VIRTIS instrument development, led by the prime contractor Leonardo-Finmeccanica (Florence, Italy), has been funded and managed by ASI, with contributions from Observatoire de Meudon financed by CNES, and from DLR. We thank the Rosetta Science Ground Segment and the Rosetta Mission Operations Centre for their support throughout all the phases of the mission. The VIRTIS calibrated data will be available through the ESA's Planetary Science Archive Website (www.rssd.esa.int/index.php?project=PSA&page=index) and is available upon request until posted to the archive.